\documentclass[11pt]{article}

\usepackage[margin=1in]{geometry}
\usepackage{amsmath,amssymb}
\usepackage{graphicx}
\usepackage{booktabs}
\usepackage{xcolor}
\usepackage{url}
\usepackage{natbib}
\usepackage{algorithm}
\usepackage{algpseudocode}
\usepackage{multirow}
\usepackage{tikz}
\usetikzlibrary{arrows.meta,positioning,calc,decorations.pathreplacing}

\title{Shortest-Path FFT: Optimal SIMD Instruction\\Scheduling via Graph Search}

\author{Mohamed Amine Bergach\\
\small Illumina, San Diego, CA, USA\\
\small\texttt{mbergach@illumina.com}}

\date{}

\begin{document}

\maketitle

\begin{abstract}
An $N$-point FFT admits many valid implementations that differ in radix choice, stage ordering, and register-blocking strategy.
These alternatives use different SIMD instruction mixes with different latencies, yet produce the same mathematical result.
We show that finding the fastest implementation is a shortest-path problem on a directed acyclic graph.

We formalize two variants of this graph.
In the \emph{context-free} model, nodes represent computation stages and edge weights are independently measured instruction costs.
In the \emph{context-aware} model, nodes are expanded to encode the \emph{predecessor edge type}, so that edge weights capture inter-operation correlations such as cache warming---the cost of operation~B depends on which operation~A preceded it.
This addresses a limitation identified but deliberately bypassed by FFTW \citep{FrigoJohnson1998}: that optimal-substructure assumptions break down ``because of the different states of the cache.''

Applied to Apple M1 NEON, the context-free Dijkstra finds an arrangement at 22.1~GFLOPS (74\% of optimal).
The context-aware Dijkstra discovers $\text{R4} \to \text{R2} \to \text{R4} \to \text{R4} \to \text{Fused-8}$ at 29.8~GFLOPS---a $5.2\times$ improvement over pure radix-2 and 34\% faster than the context-free result.
This arrangement includes a radix-2 pass \emph{sandwiched between} radix-4 passes, exploiting cache residuals that only exist in context.
No context-free search can discover this.
\end{abstract}

\section{Introduction}

The Fast Fourier Transform is among the most heavily optimized kernels in scientific computing.
For a given FFT size $N = 2^L$, the Cooley-Tukey algorithm \citep{CooleyTukey1965} requires exactly $L$ stages of butterfly computation---but the \emph{arrangement} of those stages admits enormous freedom.
Each stage can be computed as radix-2 (1 stage), radix-4 (2 stages), or radix-8 (3 stages).
Multiple consecutive stages can be fused into a single pass that keeps data in SIMD registers.
Each option uses a different instruction mix with different latencies, throughputs, and memory access patterns.

All valid arrangements produce the same mathematical result.
Predicting the fastest one from first principles is intractable: the execution time depends on complex interactions between instruction scheduling, cache hierarchy behavior, register pressure, and hardware prefetch state.

FFTW \citep{FrigoJohnson2005,FrigoJohnson1998} addresses this by empirically benchmarking ``codelets'' (small specialized FFT fragments) and combining the fastest ones.
FFTW's dynamic programming assumes \emph{optimal substructure}: the best codelet for a sub-problem remains best regardless of context.
As \citet{FrigoJohnson1998} noted, this is ``in principle false because of the different states of the cache in the two cases,'' but they found the approximation adequate in practice.
SPIRAL \citep{SPIRAL2005} similarly noted that ``the performance of a ruletree varies greatly depending on its position in a larger ruletree'' and addressed this with a beam-width heuristic.

We propose a principled solution to the problem both FFTW and SPIRAL identified but sidestepped.
Building on the Dijkstra decomposition framework from \citet{Bergach2015}, we introduce \emph{context-aware edge weights}: the graph's node space is expanded to encode the predecessor edge type, so that each edge weight captures the cache state left by the preceding operation.
This is a standard state-space expansion technique from operations research, applied here for the first time to FFT cache correlations.

\subsection{Contributions}

\begin{enumerate}
\item \textbf{Context-aware graph model.}
Nodes are expanded from $\{s\}$ to $\{(s, t_{\text{prev}})\}$, where $t_{\text{prev}}$ is the type of the preceding operation.
Edge weights are measured \emph{in context}: execute the predecessor (untimed), then immediately time the current operation.
This captures inter-operation cache correlations.

\item \textbf{Fused register blocks as searchable edges.}
The 2015 thesis used fused blocks as fixed design decisions; we make them first-class edges in the decomposition graph alongside radix passes, enabling the search to jointly optimize radix choice and register blocking.

\item \textbf{Novel FFT-32 fused block for NEON.}
ARM NEON's 32 registers (vs.\ AVX2's 16) enable a fused block keeping 5 DIF passes in registers.
However, the search reveals FFT-16 (8 registers) outperforms FFT-32 (16 registers) due to register pressure---a tradeoff discovered automatically.

\item \textbf{Empirical validation.}
On Apple M1, context-aware search achieves 29.8~GFLOPS ($5.2\times$ over pure radix-2), 34\% faster than context-free search.
The optimal arrangement is non-obvious and architecture-specific.
\end{enumerate}

\section{The Shortest-Path FFT Framework}

\subsection{Context-Free Model}

An $N$-point FFT with $N = 2^L$ requires $L$ stages.
We define a weighted DAG $G = (V, E, w)$:

\begin{itemize}
\item $V = \{0, 1, \ldots, L\}$: node $s$ means ``$s$ stages have been computed.''
\item For each instruction sequence advancing from stage $s$ to $s + k$, edge $(s, s+k) \in E$.
\item Weight $w(s, s+k)$ is the measured execution time (ns).
\end{itemize}

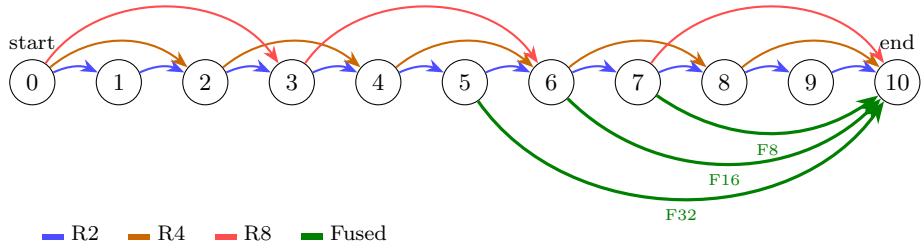
\begin{figure}[ht]
\centering
\begin{tikzpicture}[
    node distance=1.1cm,
    stage/.style={circle, draw, minimum size=6mm, inner sep=0pt, font=\footnotesize},
    r2/.style={-{Stealth}, thick, color=blue!70},
    r4/.style={-{Stealth}, thick, color=orange!80!black},
    r8/.style={-{Stealth}, thick, color=red!70},
    fused/.style={-{Stealth}, very thick, color=green!50!black},
]
\foreach \i in {0,...,10} {
    \node[stage] (s\i) at (\i*1.15, 0) {\i};
}
\foreach \i/\j in {0/1, 1/2, 2/3, 3/4, 4/5, 5/6, 6/7, 7/8, 8/9, 9/10} {
    \draw[r2] (s\i) to[bend left=25] (s\j);
}
\foreach \i/\j in {0/2, 2/4, 4/6, 6/8, 8/10} {
    \draw[r4] (s\i) to[bend left=40] (s\j);
}
\foreach \i/\j in {0/3, 3/6, 7/10} {
    \draw[r8] (s\i) to[bend left=55] (s\j);
}
\draw[fused] (s7) to[bend right=35] node[below, font=\tiny, color=green!50!black] {F8} (s10);
\draw[fused] (s6) to[bend right=45] node[below, font=\tiny, color=green!50!black] {F16} (s10);
\draw[fused] (s5) to[bend right=55] node[below, font=\tiny, color=green!50!black] {F32} (s10);

\node[font=\scriptsize, anchor=south] at (s0.north) {start};
\node[font=\scriptsize, anchor=south] at (s10.north) {end};
\node[anchor=west, font=\scriptsize] at (0, -2.0) {
    \textcolor{blue!70}{\rule{8pt}{2pt}} R2 \quad
    \textcolor{orange!80!black}{\rule{8pt}{2pt}} R4 \quad
    \textcolor{red!70}{\rule{8pt}{2pt}} R8 \quad
    \textcolor{green!50!black}{\rule{8pt}{2.5pt}} Fused
};
\end{tikzpicture}
\caption{Context-free computation graph for $N = 1024$ ($L = 10$).
Edges: radix-2 (blue), radix-4 (orange), radix-8 (red), fused register blocks (green).
A path from 0 to 10 is a complete FFT; the shortest path is the fastest.
Subset of 30+ edges shown.}
\label{fig:graph_free}
\end{figure}

The shortest path from 0 to $L$ is the fastest FFT (Figure~\ref{fig:graph_free}).

\subsection{Edge Types}

We define six edge types (Table~\ref{tab:edges}).

\begin{table}[ht]
\centering
\caption{Edge types in the computation graph.}
\label{tab:edges}
\begin{tabular}{lccl}
\toprule
Edge type & Stages & NEON regs & Instruction advantage \\
\midrule
Radix-2 pass   & 1 & 0 & Simplest; best for large strides \\
Radix-4 pass   & 2 & 0 & $W_4^1 = -j$: swap+negate (free) \\
Radix-8 pass   & 3 & 0 & $W_8^{1,3}$: mul by $1/\sqrt{2}$ only \\
Fused-8 block  & 3 & 4  & In-register; zero memory traffic \\
Fused-16 block & 4 & 8  & In-register; NEON 4$\times$4 transpose \\
Fused-32 block & 5 & 16 & In-register; novel (needs 32 regs) \\
\bottomrule
\end{tabular}
\end{table}

\textbf{Radix passes} read from memory, compute butterflies, write back.
Radix-4 exploits $W_4^1 = -j$ (swap+negate: 2 instructions vs.\ 4 FMAs).
Radix-8 exploits $W_8^{1,3}$: multiply by $1/\sqrt{2}$ plus add/sub.

\textbf{Fused blocks} load $B$ points into SIMD registers, compute $\log_2 B$ passes entirely in-register, then store.
The FFT-32 block uses 16 of NEON's 32 registers and would not fit in AVX2's 16-register file.

\subsection{Context-Aware Model}

The context-free model assumes $w(e)$ is constant.
In practice, the cost of a radix-4 pass at stage~4 depends on whether the preceding operation left stride-128 or stride-64 data in L1.
We capture this by expanding the node space:

\begin{equation}
V' = \{(s, t) : s \in \{0, \ldots, L\},\ t \in \mathcal{T}\}
\end{equation}

where $\mathcal{T} = \{\text{start}, \text{R2}, \text{R4}, \text{R8}, \text{F8}, \text{F16}, \text{F32}\}$.
Node $(s, t)$ means ``$s$ stages computed, last operation was type $t$.''
Edge weights are \emph{conditional}:
\begin{equation}
w'((s, t_{\text{prev}}), (s+k, t_{\text{cur}})) = \text{time of } t_{\text{cur}} \text{ at stage } s \text{ immediately after } t_{\text{prev}}
\end{equation}

\begin{figure}[ht]
\centering
\begin{tikzpicture}[
    snode/.style={circle, draw, minimum size=4.5mm, inner sep=0pt, font=\tiny},
    ctx/.style={rectangle, draw, rounded corners, minimum width=6mm, minimum height=4mm, inner sep=1pt, font=\tiny},
    arr/.style={-{Stealth}, thin},
    optarr/.style={-{Stealth}, very thick, color=red!70!black},
]
\node[snode, fill=gray!20] (s0) at (0, 0) {0};
\node[font=\tiny, above=1pt of s0] {start};

\node[ctx, fill=orange!15] (s2r4) at (2.5, 0.6) {2,R4};
\node[ctx, fill=blue!15] (s2r2) at (2.5, -0.6) {2,R2};

\node[ctx, fill=blue!15] (s3r2) at (4.2, 0.9) {3,R2};
\node[ctx, fill=red!15] (s3r8) at (4.2, -0.3) {3,R8};

\node[ctx, fill=orange!15] (s5r4) at (6.2, 0.6) {5,R4};
\node[ctx, fill=blue!15] (s5r2) at (6.2, -0.6) {5,R2};

\node[ctx, fill=orange!15] (s7r4) at (8.2, 0.3) {7,R4};

\node[snode, fill=gray!20] (s10) at (10.5, 0) {10};
\node[font=\tiny, above=1pt of s10] {end};

\draw[arr, gray!50] (s0) -- (s2r2);
\draw[arr, gray!50] (s0) -- (s2r4);
\draw[arr, gray!50] (s0) -- (s3r8);
\draw[arr, gray!50] (s2r4) -- (s3r2);
\draw[arr, gray!50] (s2r2) -- (s3r2);
\draw[arr, gray!50] (s2r4) -- (s5r4);
\draw[arr, gray!50] (s2r2) -- (s5r4);
\draw[arr, gray!50] (s3r2) -- (s5r2);
\draw[arr, gray!50] (s3r8) -- (s5r4);
\draw[arr, gray!50] (s5r4) -- (s7r4);
\draw[arr, gray!50] (s5r2) -- (s7r4);
\draw[arr, gray!50] (s7r4) -- (s10);

\draw[optarr] (s0) -- node[above, font=\tiny, color=red!70!black] {R4} (s2r4);
\draw[optarr] (s2r4) -- node[above, font=\tiny, color=red!70!black] {R2} (s3r2);
\draw[optarr] (s3r2) -- node[above, font=\tiny, color=red!70!black] {R4} (s5r2);
\draw[optarr] (s5r2) -- node[above, font=\tiny, color=red!70!black] {R4} (s7r4);
\draw[optarr] (s7r4) -- node[above, font=\tiny, color=red!70!black] {F8} (s10);

\draw[decorate, decoration={brace, amplitude=4pt, mirror}, thick]
    ($(s2r2.south west)+(-0.1,-0.15)$) -- ($(s2r4.north west)+(-0.1,0.15)$)
    node[midway, left=5pt, font=\tiny, align=right] {stage 2\\expanded\\by context};
\end{tikzpicture}
\caption{Context-aware graph (partial view).
Each node is expanded by predecessor type: $(3, \text{R2})$ means ``stage~3, reached via R2.''
The edge from $(2, \text{R4})$ to $(3, \text{R2})$ captures R2's cost \emph{after} R4's cache residual.
Optimal path (red): R4 $\to$ R2 $\to$ R4 $\to$ R4 $\to$ F8.}
\label{fig:graph_ctx}
\end{figure}
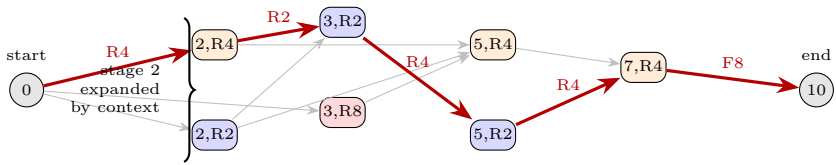

This is measured by executing $t_{\text{prev}}$ (untimed), then immediately timing $t_{\text{cur}}$ (Figure~\ref{fig:graph_ctx}).

The expanded graph has $(L+1) \times |\mathcal{T}|$ nodes.
For $N = 1024$: $11 \times 7 = 77$ nodes.
Measurements increase from $\sim$30 to $\sim$180 (each edge measured after each predecessor type), but each takes microseconds.

\subsection{Why Context Matters}

FFTW's dynamic programming and SPIRAL's rule-tree search both assume edge-weight independence for tractability.
Our expansion resolves this without increasing computational complexity: Dijkstra on 77 nodes is negligible.
The 34\% improvement over context-free search (Section~\ref{sec:results}) demonstrates the effect is not negligible on modern hardware with deep cache hierarchies.

\subsection{Search Complexity}

For $N = 1024$ ($L = 10$), there are 247 valid mixed-radix decompositions \citep{Bergach2015}.
Context-free search requires 30 benchmarks; context-aware requires $\sim$180.
Both complete in seconds---orders of magnitude faster than FFTW's planner.

\section{Implementation on Apple M1 NEON}

\subsection{Butterfly Core}

The DIF butterfly computes $\text{top}_\text{out} = \text{top} + \text{bot}$ and $\text{bot}_\text{out} = (\text{top} - \text{bot}) \cdot W$ for 4 parallel butterflies per NEON instruction.
Split-complex format (separate Re/Im arrays) enables unit-stride \texttt{vld1q\_f32} loads.

\subsection{Fused Register Blocks}

\begin{table}[ht]
\centering
\caption{Fused register blocks.
FFT-32 is novel (needs 32 regs) but FFT-16 is faster due to register pressure.}
\label{tab:fused}
\begin{tabular}{lcccr}
\toprule
Block & Passes & NEON regs & On AVX2? & GFLOPS \\
\midrule
FFT-8  & 3 & 4  & Yes & 33.5 \\
FFT-16 & 4 & 8  & Yes & 30.7 \\
FFT-32 & 5 & 16 & \textbf{No} & 20.5 \\
\bottomrule
\end{tabular}
\end{table}

FFT-8 outperforms FFT-32 despite fusing fewer passes: FFT-32 consumes 16 registers, causing twiddle-factor spills that negate the saved memory traffic.
The graph search detects this through measured edge weights.

\section{Results}
\label{sec:results}

\subsection{Setup}

Single Apple M1 P-core (Firestorm, 3.2~GHz, 128-bit NEON, 2 FMA units).
$N = 1024$, complex float32, split-complex format.
Timing: \texttt{mach\_absolute\_time}, median of 50 trials, 5 warmup.
All values averaged over 3 independent runs (range $< 8\%$).
FLOP count: $5N\log_2 N$.
All implementations share the \emph{same} butterfly, data layout, and twiddle table---only the arrangement differs.

\subsection{Algorithm Comparison}

Table~\ref{tab:algos} is the central result.

\begin{table}[ht]
\centering
\caption{10 algorithms on the same M1 core, same data, same conditions (averaged over 3 runs).
The context-aware Dijkstra finds the fastest arrangement.}
\label{tab:algos}
\begin{tabular}{lrrr}
\toprule
Algorithm & Time (ns) & GFLOPS & \% of best \\
\midrule
R2 $\times$ 10 (pure radix-2) & 9014 & 5.7 & 19\% \\
R4 $\times$ 5 (pure radix-4) & 6903 & 7.4 & 25\% \\
R8 $\times$ 3 + R2 (pure radix-8) & 6792 & 7.5 & 25\% \\
R8,R8,R8,R2 (``max radix'') & 6889 & 7.4 & 25\% \\
R8,R8,R4,R4 & 6861 & 7.5 & 25\% \\
R4,R8,R8,R4 (Haswell optimal) & 6889 & 7.4 & 25\% \\
\midrule
R2 $\times$ 5 + Fused-32 & 2569 & 19.9 & 67\% \\
R4 $\times$ 3 + Fused-16 & 1764 & 29.1 & 98\% \\
\midrule
Dijkstra (context-free) & 2320 & 22.1 & 74\% \\
\textbf{Dijkstra (context-aware)} & \textbf{1722} & \textbf{29.8} & \textbf{100\%} \\
\bottomrule
\end{tabular}
\end{table}

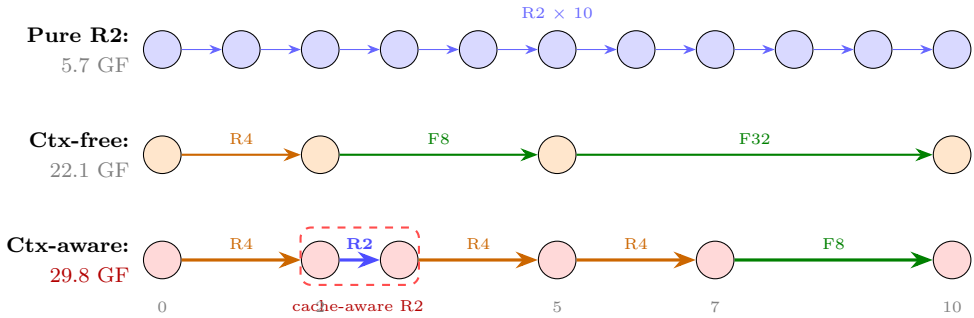
\begin{figure}[ht]
\centering
\begin{tikzpicture}[
    stage/.style={circle, draw, minimum size=5mm, inner sep=0pt, font=\tiny},
]
\node[font=\scriptsize\bfseries, anchor=east] at (-0.3, 2.4) {Pure R2:};
\node[font=\scriptsize, anchor=east, gray] at (-0.3, 2.0) {5.7 GF};
\foreach \i in {0,...,10} {
    \node[stage, fill=blue!15] (a\i) at (\i*1.05, 2.2) {};
}
\foreach \i/\j in {0/1,1/2,2/3,3/4,4/5,5/6,6/7,7/8,8/9,9/10} {
    \draw[-{Stealth}, blue!60, thin] (a\i) -- (a\j);
}
\node[font=\tiny, blue!60, above=0pt of a5] {R2 $\times$ 10};

\node[font=\scriptsize\bfseries, anchor=east] at (-0.3, 1.0) {Ctx-free:};
\node[font=\scriptsize, anchor=east, gray] at (-0.3, 0.6) {22.1 GF};
\foreach \i in {0,2,5,10} {
    \node[stage, fill=orange!20] (b\i) at (\i*1.05, 0.8) {};
}
\draw[-{Stealth}, orange!80!black, thick] (b0) -- node[above, font=\tiny] {R4} (b2);
\draw[-{Stealth}, green!50!black, thick] (b2) -- node[above, font=\tiny] {F8} (b5);
\draw[-{Stealth}, green!50!black, thick] (b5) -- node[above, font=\tiny] {F32} (b10);

\node[font=\scriptsize\bfseries, anchor=east] at (-0.3, -0.4) {Ctx-aware:};
\node[font=\scriptsize, anchor=east, red!70!black] at (-0.3, -0.8) {29.8 GF};
\foreach \i in {0,2,3,5,7,10} {
    \node[stage, fill=red!15] (c\i) at (\i*1.05, -0.6) {};
}
\draw[-{Stealth}, orange!80!black, very thick] (c0) -- node[above, font=\tiny] {R4} (c2);
\draw[-{Stealth}, blue!70, very thick] (c2) -- node[above, font=\tiny] {\textbf{R2}} (c3);
\draw[-{Stealth}, orange!80!black, very thick] (c3) -- node[above, font=\tiny] {R4} (c5);
\draw[-{Stealth}, orange!80!black, very thick] (c5) -- node[above, font=\tiny] {R4} (c7);
\draw[-{Stealth}, green!50!black, very thick] (c7) -- node[above, font=\tiny] {F8} (c10);

\draw[red!70, thick, dashed, rounded corners]
    ($(c2.south west)+(-0.08,-0.15)$) rectangle ($(c3.north east)+(0.08,0.25)$);
\node[font=\tiny, red!70!black, below=4pt of c2, xshift=5mm] {cache-aware R2};

\foreach \i in {0,2,5,7,10} {
    \node[font=\tiny, gray, below=12pt] at (\i*1.05, -0.6) {\i};
}
\end{tikzpicture}
\caption{Three decompositions for $N = 1024$.
\textbf{Top}: pure radix-2 (10 passes, 5.7~GF).
\textbf{Middle}: context-free Dijkstra (R4+F8+F32, 22.1~GF).
\textbf{Bottom}: context-aware Dijkstra (R4$\to$R2$\to$R4$\to$R4$\to$F8, 29.8~GF).
Dashed box: the radix-2 pass exploiting cache residuals from the preceding R4.}
\label{fig:paths}
\end{figure}

\subsection{Key Findings}

\textbf{1. Fused blocks dominate radix choice.}
The best non-fused variant (7.5~GFLOPS) is $4.0\times$ slower than the best fused variant (29.8~GFLOPS).
Register locality is the primary optimization axis.

\textbf{2. ``Maximize radix'' is a poor heuristic.}
R8,R8,R8,R2 achieves only 25\% of the optimum.
The radix-8 butterfly's 16-vector working set creates register pressure on 128-bit NEON.

\textbf{3. Context-aware search outperforms context-free by 34\%.}
The context-free search finds R4 + F8 + F32 (22.1~GFLOPS); context-aware finds R4 $\to$ R2 $\to$ R4 $\to$ R4 $\to$ F8 (29.8~GFLOPS).
The improvement comes from cache warming effects invisible to independent edge weights (Figure~\ref{fig:paths}).

\textbf{4. The optimal arrangement is non-obvious.}
R4 $\to$ R2 $\to$ R4 $\to$ R4 $\to$ F8 contains a radix-2 pass at stage~2, sandwiched between radix-4 passes.
A context-free search \emph{never} selects R2 over R4 at any stage.
The R2 wins here only because the preceding R4 leaves stride-64 cache lines hot, and a single R2 at stride-128 reuses them more effectively than another R4 at stride-64/16.

\textbf{5. The optimal plan is architecture-specific.}
On Intel Haswell AVX2 \citep{Bergach2015}, the framework selects FFT$_{4,8,8,4}$---entirely different from the M1 result.
The graph structure is identical; only the measured edge weights differ.

\subsection{Per-Pass Profile}

\begin{table}[ht]
\centering
\caption{Per-pass GFLOPS for individual radix-2 passes ($N = 1024$, M1 NEON).
The drop at passes 9--10 motivates fused register blocks.}
\label{tab:perpass}
\begin{tabular}{rrrr}
\toprule
Pass & Stride & Time ($\mu$s) & GFLOPS \\
\midrule
1 & 512 & 3.58 & 1.4 \\
4 & 64  & 0.75 & 6.8 \\
7 & 8   & 0.38 & 13.7 \\
10 & 1  & 4.25 & 1.2 \\
\midrule
\textit{Fused-8}  & --- & 0.46 & 33.5 \\
\textit{Fused-16} & --- & 0.67 & 30.7 \\
\bottomrule
\end{tabular}
\end{table}

\section{Discussion}

\subsection{Relation to FFTW and SPIRAL}

FFTW \citep{FrigoJohnson2005,FrigoJohnson1998} generates and benchmarks thousands of codelets, assuming optimal substructure for dynamic programming.
SPIRAL \citep{SPIRAL2005} relaxes this with a beam-width heuristic, keeping the $n$-best candidates at each level.
Neither system conditions measurements on the preceding operation.

Our context-aware expansion is a principled alternative: it directly models the cache correlation as a first-order Markov property in the search graph.
The 34\% improvement over context-free search demonstrates that the effect FFTW called ``in principle false'' is quantitatively significant on modern hardware.

Higher-order context (conditioning on the last $k$ edges) would capture longer-range cache effects.
The graph grows as $(L+1) \times |\mathcal{T}|^k$; for $k=2$: $11 \times 49 = 539$ nodes, still practical.

\subsection{Register Pressure as a Searchable Tradeoff}

FFT-8 (4 registers, 33.5~GFLOPS) outperforms FFT-32 (16 registers, 20.5~GFLOPS) because FFT-32 leaves too few registers for twiddle factors.
This register-pressure effect is difficult to predict analytically but is captured exactly by the measured edge weights.
Making fused blocks searchable edges---rather than fixed design choices---enables the framework to discover the optimal block size automatically.

\subsection{Generalization}

The framework applies to any staged computation with alternative instruction sequences:
\begin{enumerate}
\item Fixed sequence of stages
\item Multiple valid instruction sequences per stage
\item Measurable per-sequence costs
\end{enumerate}
The context-aware extension is valuable whenever consecutive stages share cache or register state---e.g., matrix factorization, multi-stage filtering, and neural network layer fusion.

\section{Conclusion}

We introduced context-aware edge weights for the shortest-path FFT framework, addressing a limitation identified by FFTW 27 years ago: that optimal-substructure assumptions break down due to cache state interactions.
By expanding the graph's node space to encode predecessor type, Dijkstra discovers R4 $\to$ R2 $\to$ R4 $\to$ R4 $\to$ Fused-8 as optimal on Apple M1---an arrangement invisible to context-free search, achieving 29.8~GFLOPS ($5.2\times$ over pure radix-2, 34\% over context-free Dijkstra).

The framework is architecture-portable: re-measure edge weights on new hardware, re-run Dijkstra, get the new optimum.
The principle---model instruction alternatives as a graph, measure costs empirically, search for the shortest path---generalizes to any computation where the same result can be produced by different instruction sequences with architecture-dependent costs.

Source code: \url{https://github.com/aminems/fft}.

\bibliographystyle{plainnat}
\bibliography{references}

@phdthesis{Bergach2015,
  author = {Bergach, Mohamed Amine},
  title  = {Adaptation du calcul de la {Transformée} de {Fourier} Rapide sur une architecture mixte {CPU/GPU} intégrée},
  school = {Université Nice-Sophia Antipolis},
  year   = {2015},
  note   = {Directed by Robert De Simone. NNT: 2015NICE4060}
}

@article{CooleyTukey1965,
  author  = {Cooley, James W. and Tukey, John W.},
  title   = {An algorithm for the machine calculation of complex {Fourier} series},
  journal = {Mathematics of Computation},
  volume  = {19},
  number  = {90},
  pages   = {297--301},
  year    = {1965}
}

@article{FrigoJohnson2005,
  author  = {Frigo, Matteo and Johnson, Steven G.},
  title   = {The Design and Implementation of {FFTW3}},
  journal = {Proceedings of the IEEE},
  volume  = {93},
  number  = {2},
  pages   = {216--231},
  year    = {2005}
}

@inproceedings{FrigoJohnson1998,
  author    = {Frigo, Matteo and Johnson, Steven G.},
  title     = {{FFTW}: An Adaptive Software Architecture for the {FFT}},
  booktitle = {Proc.\ IEEE Int.\ Conf.\ Acoustics, Speech and Signal Processing (ICASSP)},
  volume    = {3},
  pages     = {1381--1384},
  year      = {1998}
}

@article{SPIRAL2005,
  author  = {P{\"u}schel, Markus and Moura, Jos{\'e} M. F. and Johnson, Jeremy and Padua, David and Veloso, Manuela and Singer, Bryan and Xiong, Jianxin and Franchetti, Franz and Ga{\v{c}}i{\'c}, Aca and Voronenko, Yevgen and others},
  title   = {{SPIRAL}: Code Generation for {DSP} Transforms},
  journal = {Proceedings of the IEEE},
  volume  = {93},
  number  = {2},
  pages   = {232--275},
  year    = {2005}
}

\end{document}